\documentclass[a4paper,11pt]{article}
\pdfoutput=1 % if your are submitting a pdflatex (i.e. if you have
\usepackage{jheppub} % for details on the use of the package, please
\usepackage[T1]{fontenc} % if needed
\usepackage{xcolor} % Add this line
\usepackage{hyperref} % You can still use hyperref for other options

\title{\boldmath Gravitational Wave Constraints on the Bouncing Energy Scale of Big Bounce Cosmology}

\author[a]{Changhong Li}
\emailAdd{changhongli@ynu.edu.cn}

\affiliation[a]{Department of Astronomy, Key Laboratory of Astroparticle Physics of Yunnan Province,  School of Physics and Astronomy,  Yunnan University, No.2 Cuihu North Road, Kunming,  650091 China}

\abstract{Big bounce cosmology provides a solution to the Universe's initial singularity, and stochastic gravitational wave background (SGWB) searches offer a promising avenue for testing this paradigm. In this work, we establish an analytical relation between the bouncing energy scale, $\rho_{s\downarrow}^{1/4}$, and SGWB spectrum, $\Omega_\mathrm{GW}(f)h^2$, for big bounce cosmology. By combining sensitivities from major GW detectors (e.g., Planck/BICEP, PTA, and LIGO/Virgo across low, medium, and high frequencies, respectively), we provide the first systematic GW constraint on $\rho_{s\downarrow}^{1/4}$. Our results show that the region $-\tfrac{1}{3} < w_1 < -0.17$ is excluded by current SGWB searches, given the constraint $\rho_{s\downarrow}^{1/4} > 1~\mathrm{TeV}$, where $w_1$ is the contraction-phase equation of state parameter. Additionally, no detectable SGWB can be generated for $0.038 < w_1 < \infty$ with $\rho_{s\downarrow}^{1/4} < 10^{16}~\mathrm{TeV}$. We identify a window, $-0.17 < w_1 < 0.038$, in which a detectable SGWB can be produced, disfavoring nearly all big bounce models except for the matter-dominated contraction model ($w_1 \simeq 0$).}

\begin{document}
\maketitle
\flushbottom

%%%%%%%%%%%%%%%%%%%%%%%%%%%%%%%%%%%%%%%%%%%%%%%%%%%%%
\section{Introduction}

Big bounce cosmology provides a compelling solution to the Universe's initial singularity problem, proposing that the Universe transitions from contraction to expansion with a non-zero minimal size during the bounce phase~\cite{Khoury:2001wf, Gasperini:2002bn, Creminelli:2006xe, Peter:2006hx, Cai:2007qw, Cai:2008qw, Saidov:2010wx, Li:2011nj, Cai:2011tc, Easson:2011zy, Bhattacharya:2013ut, Qiu:2015nha, Cai:2016hea, Barrow:2017yqt, deHaro:2017yll, Ijjas:2018qbo, Boruah:2018pvq, Nojiri:2019yzg, Silva:2015qna, Silva:2020bnn,Silva:2023ieb, Nayeri:2005ck, Brandenberger:2006xi, Fischler:1998st, Cai:2009rd, Li:2014era, Cheung:2014nxi, Li:2014cba, Li:2015egy, Li:2020nah} (for comprehensive reviews, see~\cite{Novello:2008ra, Brandenberger:2016vhg, Nojiri:2017ncd, Odintsov:2023weg} and references therein). This paradigm contrasts with standard inflationary cosmology~\cite{Guth:1980zm, Starobinsky:1980te, Sato:1980yn, Linde:1981mu, Albrecht:1982wi, Mukhanov:1990me}, which assumes a singular beginning~\cite{Borde:1993xh, Borde:2001nh}. Current and upcoming stochastic gravitational wave background (SGWB) searches offer a promising means to test the big bounce model~\cite{Caprini:2018mtu} (also see \cite{Li:2024oru}, and references therein). However, compared to the simpler inflationary scenario, the non-singular structure of big bounce cosmology presents a more intricate framework for investigating the generation and evolution of primordial gravitational waves (PGWs) (see \cite{Li:2024dce}, and references therein). Specifically, in big bounce cosmology, PGWs exit and reenter the horizon twice (see Fig.~\ref{fig:HRinBounce.pdf}), whereas in standard inflation, this process occurs only once.

Recently, in Ref.~\cite{Li:2024dce}, we addressed the challenge of computing the PGW spectrum, $ \mathcal{P}_h(k) $, in big bounce cosmology by deriving an abstract matrix representation for $ \mathcal{P}_h(k) $ applicable to a general big bounce cosmology (see Eq.~(\ref{eq:phfpketa}) and Eqs.~(\ref{eq:nxxdefine})–(\ref{eq:m3dlemexp})). This was achieved by explicitly solving the equation of motion for PGWs and matching solutions at the phase boundaries. However, the resulting general mathematical expression cannot be directly compared with astrophysical observations from gravitational wave (GW) searches. 

In this work, we derive an explicit expression for the PGW spectrum, $ \mathcal{P}_h(k) $, from the aforementioned abstract matrix representation, by incorporating three physical conditions that apply to a realistic realization of big bounce cosmology. First, we assume that the Universe is radiation-dominated after the bounce, in accordance with expectations for a standard post-reheating radiation era. Second, we assume that the region near the bounce point is dominated by an exotic form of matter with a highly negative equation of state (e.g., Quintom matter~\cite{Cai:2009zp}), ensuring the occurrence of the bounce. Third, we assume the bounce process around the bounce point is symmetric for simplicity.

With this explicit expression for $ \mathcal{P}_h(k) $, we establish, for the first time, an analytical relation between the bouncing energy scale of big bounce cosmology, $ \rho_{s\downarrow}^{1/4} $, and the SGWB spectrum, $ \Omega_\mathrm{GW}(f)h^2 $. The bouncing energy scale $ \rho_{s\downarrow}^{1/4} $ represents the quasi-highest energy scale of big bounce cosmology, evaluated at the beginning of the post-bounce expansion (see Eq.~(\ref{eq:rhosdd})). This relation provides a means to probe the energy scale and spatial extent of the early non-singular Universe, while also uncovering new physics through gravitational wave searches.

By combining sensitivities from major GW detectors (see ~\cite{Schmitz:2020syl, Annis:2022xgg, Bi:2023tib} and references therein), e.g., Planck/BICEP, PTA(NanoGrav, PPTA, EPTA), and aLIGO + aVirgo (O2) at low, medium, and high frequencies, respectively, with this analytical framework, we provide the first systematic gravitational wave constraint on $ \rho_{s\downarrow}^{1/4} $ (see Fig.~\ref{fig:Fullplotasrhos.pdf} and Fig.~\ref{fig:HighResw1asrho.pdf}). Our results show that the region $ -\tfrac{1}{3} < w_1 < -0.17 $ is excluded by current SGWB searches, given the physical constraint $ \rho_{s\downarrow}^{1/4} > 1~\mathrm{TeV} $, where $ w_1 $ is the contraction-phase equation of state parameter. Furthermore, we find that no detectable SGWB can be generated for $ 0.038 < w_1 < \infty $ with $ \rho_{s\downarrow}^{1/4} < 10^{16}~\mathrm{TeV} $. We identify a window, $ -0.17 < w_1 < 0.038 $, in which a detectable SGWB can be produced within the physically acceptable range of $ 1~\mathrm{TeV} < \rho_{s\downarrow}^{1/4} < 10^{16}~\mathrm{TeV} $. This window excludes nearly all big bounce models except those featuring a nearly matter-dominated contraction ($ w_1 \simeq 0 $), where $ w_1 = 0 $ corresponds to the scale-invariant solution in the parameter space for $ \mathcal{P}_h(k) $ (and $ \Omega_\mathrm{GW}(f) h^2 $).

Additionally, we find, for $ w_1 < 0 $, the improvement from CMB-S4 is moderate in the exclusion region in parameter space $ (w_1, \rho_{s\downarrow}) $, yet potentially decisive: a future CMB-S4 detection of gravitational waves would confirm whether such a signal originates from PGWs in big bounce cosmology. If confirmed, both $ w_1 $ and $ \rho_{s\downarrow}^{1/4} $ can be determined (see Eq.~(\ref{eq:OmegaGWCnrhos})). On the other hand, for $ w_1 > 0 $, although upcoming detectors (e.g., DECIGO, BBO, LISA, Taiji, TianQin, Cosmic Explorer, and the Einstein Telescope) will significantly improve current limits, they will still fall short of the theoretical constraint at the Planck scale. This highlights a strong theoretical motivation for the construction of new high-frequency GW detectors (i.e., $ f \gg 100~\mathrm{Hz} $) in the future.

%%%%%%%%%%%%%%%%%%%%%%%%%%%%%%%%%%%%%%%%%%%%%%%%%%%%%
\section{Stochastic Gravitational Waves for Big Bounce Cosmology}
The spectrum of the SGWB induced by PGWs is given by \cite{Caprini:2018mtu}:
\begin{equation} \label{eq:sgwbphf}
   \Omega_\mathrm{GW}(f)h^2 = \frac{1}{24}\Omega_{\gamma 0} h^2 \cdot \mathcal{P}_h(f) \mathcal{T}_{\mathrm{eq}}(f).
\end{equation}
Here, $ f = k / (2\pi a_0) $ is the observed frequency today ($ a_0 = 1 $), where $ k $ is the comoving wavevector of PGWs. The term $ \mathcal{P}_h(f) = \mathcal{P}_h(\eta_k)|_{k = 2\pi f a_0} $ represents the PGW spectrum at horizon reentry ($ k\eta_k = 1 $) during the radiation-dominated era, while $ \mathcal{T}_{\mathrm{eq}}(f) \equiv \left[1 + \frac{9}{32} \left( f_\mathrm{eq}/f \right)^2 \right] $ accounts for the difference between modes reentering the horizon during the radiation-dominated and matter-dominated eras. Here, $ f_\mathrm{eq} = 2.01 \times 10^{-17}~\mathrm{Hz} $ is the frequency at matter-radiation equality. Additionally, $ \Omega_{\gamma 0} h^2 = 2.474 \times 10^{-5} $ represents the energy density fraction of radiation today, with $ h = 0.677 $ being the reduced Hubble constant.

\begin{figure}[htbp]
\centering 
\includegraphics[width=0.8\textwidth]{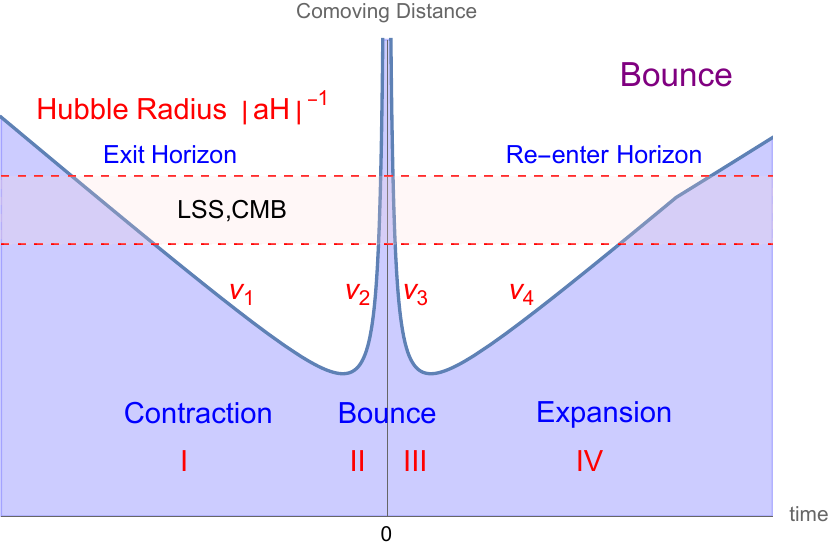}
\caption{\label{fig:HRinBounce.pdf} Illustration of the cosmic evolution of the effective Hubble radius ($ |aH|^{-1} $) within the framework of big bounce cosmology. The horizon shrinks during Phase I and Phase III, and expands during Phase II, and Phase IV. The PGWs relevant for SGWB searches are generated during Phase I, propagate through Phases II and III, and reenter the horizon during and after Phase IV.}
\end{figure}

In big bounce cosmology, the cosmic evolution prior to matter-radiation equality involves four distinct phases: Phase I — collapsing contraction ($\dot{a}<0$ and $\ddot{a}<0$), Phase II — bouncing contraction ($\dot{a}<0$ and $\ddot{a}>0$), Phase III — bouncing expansion ($\dot{a}>0$ and $\ddot{a}>0$), and Phase IV — decelerating expansion ($\dot{a}>0$ and $\ddot{a}<0$), labeled by $i = 1, 2, 3, 4$ respectively (see Fig.~\ref{fig:HRinBounce.pdf}), where $\dot{a}$ and $\ddot{a}$ respectively represent the velocity and acceleration of the scale factor of the Universe, $a$, in terms of physical time, $t$~\cite{Li:2024dce}. In each phase, assuming a constant equation of state (EoS) for the cosmic background $w_i$, the scale factor $a(\eta) = a_i |\eta|^{\nu_i}$ evolves according to the power-law index $\nu_i$, where $\eta$ is the conformal time. The relationship between $\nu_i$ and $w_i$ is given by $\nu_i = \frac{2}{3w_i + 1}$. The boundaries of each phase, $\eta_{i\downarrow/\uparrow}$, represent the transition points between phases ($i$ and $i+1$), with the subscripts $\downarrow$ and $\uparrow$ indicating the superhorizon ($k\eta_{i\downarrow} \ll 1$) and  subhorizon($k\eta_{i\uparrow} \gg 1$) regimes, respectively.

In Fig.~\ref{fig:HRinBounce.pdf}, following~\cite{Cheung:2016vze, Li:2024dce}, we illustrate the cosmic evolution of the effective Hubble radius $\left( |aH|^{-1} \right)$ within the framework of general big bounce cosmology. This illustration shows that the horizon shrinks during Phase I — collapsing contraction and Phase III — bouncing expansion, and expands during Phase II — bouncing contraction, and Phase IV — decelerating expansion. The PGWs, relevant for SGWB searches, are generated during Phase I, propagate through Phases II and III, and reenter the horizon during and after Phase IV.

In Ref.~\cite{Li:2024dce}, by explicitly solving the equation of motion for tensor perturbations in each of the four phases, matching solutions at the boundaries from Phase I to Phase IV, we derived an abstract matrix representation of the PGW spectrum, $ \mathcal{P}_h(k) $, for a general big bounce cosmology, at the horizon reentry ($ \eta_k $) during Phase IV:
\begin{equation}\label{eq:phfpketa}
    \mathcal{P}_h(\eta_k) \equiv \frac{k^3(|h_{k+}^{[4]}|^2 + |h_{k \times}^{[4]}|^2)}{2\pi^2}
    = \frac{(k\eta_k)^3}{2\pi^2} \frac{\pi}{\nu_4^2} \frac{[H(\eta_k)]^2}{m_\mathrm{pl}^2} N_{22}(\{\Tilde{\nu}_i\}, \{\eta_{i\downarrow/\uparrow}\})~,
\end{equation}
where $h_{k+/\times}^{[4]}$ are Fourier modes of the two polarization states of transverse and traceless tensor perturbations, $ \{\Tilde{\nu}_i\} = (\Tilde{\nu}_1, \Tilde{\nu}_2, \Tilde{\nu}_3, \Tilde{\nu}_4) $ with $\Tilde{\nu}_i \equiv |\nu_i - \tfrac{1}{2}|$ being the reduced power-law indices for each phase, and $\{\eta_{i\downarrow/\uparrow}\} = (\eta_{1\downarrow}, \eta_{2\uparrow}, \eta_{3\downarrow})$ representing the boundaries between phases. $H(\eta_k)$ is the Hubble parameter evaluated at horizon reentry $\eta=\eta_k$, and the reduced Planck mass is denoted as $m_\mathrm{pl} \equiv \sqrt{8\pi G}^{-1}$. The term $N_{22}(\{\Tilde{\nu}_i\}, \{\eta_{i\downarrow/\uparrow}\})$ represents the $_{22}$-component of the propagation matrix of PGWs, $N(\{\Tilde{\nu}_i\}, \{\eta_{i\downarrow/\uparrow}\})$, which is constructed from the transformation matrices $T_i$ (or their inverses $T_i^{-1}$) for each phase, along with the matching matrices $M_{i\uparrow/\downarrow}$ for the phase boundaries: 
\begin{equation}\label{eq:nxxdefine}
    N(\{\Tilde{\nu}_i\}, \{\eta_{i\downarrow/\uparrow}\}) \equiv X^\dagger X~,\quad 
\end{equation}
where 
\begin{equation}
    X \equiv T_4^{-1} M_{3\downarrow} T_3 M_{2\uparrow} T_2^{-1} M_{1\downarrow} T_1~,
\end{equation}
and 
\begin{equation}
    T_i = 
    \begin{pmatrix}
        \alpha_i & \alpha_i^\ast \\
        \beta_i & \beta_i^\ast
    \end{pmatrix}~,
\end{equation}
where 
\begin{equation}
    \alpha_i \equiv 2^{-\Tilde{\nu}_i} \left[ \frac{-i e^{i\pi \Tilde{\nu}_i}}{\sin(\pi \Tilde{\nu}_i) \Gamma(\Tilde{\nu}_i + 1)} \right]~, \quad 
    \beta_i \equiv -i \left[ \frac{\Gamma(\Tilde{\nu}_i)}{\pi} 2^{\Tilde{\nu}_i} \right]~,
\end{equation}
and $\Gamma(z) = \int_0^\infty t^{z-1} e^{-t} \, dt$ is the Gamma function. Additionally,
\begin{equation}\label{eq:m2uexp}
    M_{2\uparrow}
    = \begin{pmatrix}
        e^{-i(\Tilde{\nu}_2 - \Tilde{\nu}_3)\pi/2} & 0 \\
        0 & e^{i(\Tilde{\nu}_2 - \Tilde{\nu}_3)\pi/2}
    \end{pmatrix}~,
\end{equation}

\begin{equation} \label{eq:mnu1gt}
    M_{1\downarrow > \tfrac{1}{2}} =
    \begin{pmatrix}
        0 & -\frac{\Tilde{\nu}_1}{\Tilde{\nu}_2} \left(k \eta_{1\downarrow}\right)^{-\left(\Tilde{\nu}_1 + \Tilde{\nu}_2\right)} \\
        \left(k \eta_{1\downarrow}\right)^{\Tilde{\nu}_1 + \Tilde{\nu}_2} & \left(1 + \frac{\Tilde{\nu}_1}{\Tilde{\nu}_2}\right) \left(k \eta_{1\downarrow}\right)^{-\left(\Tilde{\nu}_1 - \Tilde{\nu}_2\right)}
    \end{pmatrix}~,\quad \nu_1 > \tfrac{1}{2}~,
\end{equation}

\begin{equation}\label{eq:mnu1le}
    M_{1\downarrow \le \tfrac{1}{2}} =
    \begin{pmatrix}
        \frac{\Tilde{\nu}_1}{\Tilde{\nu}_2} \left(k \eta_{1\downarrow}\right)^{\Tilde{\nu}_1 - \Tilde{\nu}_2} & 0 \\
        \left(1 - \frac{\Tilde{\nu}_1}{\Tilde{\nu}_2}\right) \left(k \eta_{1\downarrow}\right)^{\Tilde{\nu}_1 + \Tilde{\nu}_2} & \left(k \eta_{1\downarrow}\right)^{-\left(\Tilde{\nu}_1 - \Tilde{\nu}_2\right)}
    \end{pmatrix}~,\quad \nu_1 \le \tfrac{1}{2}~,
\end{equation}

\begin{equation}\label{eq:m3dgtmexp}
    M_{3\downarrow > \tfrac{1}{2}} =
    \begin{pmatrix}
        \left(1 + \frac{\Tilde{\nu}_3}{\Tilde{\nu}_4}\right) \left(k \eta_{3\downarrow}\right)^{\Tilde{\nu}_3 - \Tilde{\nu}_4} & \left(k \eta_{3\downarrow}\right)^{-\left(\Tilde{\nu}_3 + \Tilde{\nu}_4\right)} \\
        -\frac{\Tilde{\nu}_3}{\Tilde{\nu}_4} \left(k \eta_{3\downarrow}\right)^{\Tilde{\nu}_3 + \Tilde{\nu}_4} & 0
    \end{pmatrix}~,\quad \nu_4 > \tfrac{1}{2}~.
\end{equation}

\begin{equation}\label{eq:m3dlemexp}
    M_{3\downarrow \le \tfrac{1}{2}} =
    \begin{pmatrix}
        \frac{\Tilde{\nu}_3}{\Tilde{\nu}_4} \left(k \eta_{3\downarrow}\right)^{\Tilde{\nu}_3 - \Tilde{\nu}_4} & 0 \\
        \left(1 - \frac{\Tilde{\nu}_3}{\Tilde{\nu}_4}\right) \left(k \eta_{3\downarrow}\right)^{\Tilde{\nu}_3 + \Tilde{\nu}_4} & \left(k \eta_{3\downarrow}\right)^{-\left(\Tilde{\nu}_3 - \Tilde{\nu}_4\right)}
    \end{pmatrix}~,\quad \nu_4 \le \tfrac{1}{2}~,
\end{equation}
In Eq.~(\ref{eq:m2uexp}), $ M_{2\uparrow} $ takes a single expression as $ \nu_2 < 0 < \frac{1}{2} $ and $ \nu_3 < 0 < \frac{1}{2} $ for Phase II and Phase III. For a more detailed interpretation of these matrices, see Ref.~\cite{Li:2024dce}.

In this work, to impose observational constraints from gravitational waves on big bounce cosmology, we apply the following three physical conditions for a realistic realization of big bounce cosmology to compute $ N_{22}(\{\Tilde{\nu}_i\}, \{\eta_{i\downarrow/\uparrow}\}) $ explicitly. First, we assume that the Universe is radiation-dominated in Phase IV ($ w_4 = \frac{1}{3} $), which corresponds to $ \nu_4 = 1 $, as expected for a standard post-reheating radiation era. Second, we assume that the bouncing region near the bounce point is dominated by an exotic form of matter with a highly negative equation of state (e.g., Quintom matter~\cite{Cai:2009zp}), ensuring the occurrence of the bounce, i.e., $ w_2 = w_3 = -\infty $. Third, we assume a symmetric bounce, i.e., $ \eta_{s\downarrow} \equiv \eta_{1\downarrow} = \eta_{3\downarrow} $.

Under these three assumptions, our model reduces to:
\begin{equation}\label{eq:ourmodelw}
        \{w_i\}=(w_1, w_2, w_3, w_4) = \left(w_1, -\infty, -\infty, \tfrac{1}{3}\right)~,
\end{equation}
and
\begin{equation}\label{eq:ourmodeleta}
        \{\eta_{i\downarrow/\uparrow}\}=(\eta_{1\downarrow},\eta_{2\uparrow},\eta_{3\downarrow})=(\eta_{s\downarrow},\infty,\eta_{s\downarrow})~,
\end{equation}
where $ \eta_{2\uparrow} \rightarrow \infty $ because it is the bouncing point ($ H(\eta_{2\uparrow}) = 0 $), and $ w_1 \in \left[-\tfrac{1}{3}, \infty\right) $ for the collapsing contraction phase ($ \dot{a} < 0 $ and $ \ddot{a} < 0 $)~\cite{Li:2024dce}. Additionally, using \begin{equation}\label{eq:nutnuwrelation}
    \nu_i=\frac{2}{3w_i+1} \quad\text{and}\quad \Tilde{\nu}_i\equiv |\nu_i - \tfrac{1}{2}|=\left|\frac{3(1-w_i)}{2(3w_i+1)}\right|~,
\end{equation}
we obtain
\begin{equation} \label{eq:ourmodelnu}
        \{\nu_i\}=(\nu_1,\nu_2,\nu_3,\nu_4)=(\nu_1,0,0,1)~,
\end{equation}
and
\begin{equation} \label{eq:ourmodeltnu}
        \{ \Tilde{\nu}_i\}=(\Tilde{\nu}_1,\Tilde{\nu}_2,\Tilde{\nu}_3,\Tilde{\nu}_4)=\left(\Tilde{\nu}_1,\tfrac{1}{2},\tfrac{1}{2},\tfrac{1}{2}\right)~.
\end{equation}
Substituting Eq.(\ref{eq:ourmodeleta}) and Eq.(\ref{eq:ourmodeltnu}) into Eq.(\ref{eq:nxxdefine})-Eq.(\ref{eq:m3dgtmexp}) and using Eq.(\ref{eq:nutnuwrelation}) again, we obtain the leading order term of $ N_{22}(\{\Tilde{\nu}_i\}, \{\eta_{i\downarrow/\uparrow}\})$ in the deep bouncing limit $k\eta_{s\downarrow}\ll 1$, 
\begin{equation} \label{eq:NCkesnw1}
    N_{22}(w_1) = C(w_1) \frac{1}{(k\eta_{s\downarrow})^{n(w_1)}}~,
\end{equation}
where
\begin{equation}\label{eq:Cw1}
    C(w_1) = 
\begin{cases} 
    \pi^{-1}4^{-1-\frac{3(1-w_1)}{2(3w_1+1)}}\Gamma^2\left(-\frac{3(1-w_1)}{2(3w_1+1)}\right), & \text{if } w_1\ge1 \\
    \pi^{-1}4^{-1+\frac{3(1-w_1)}{2(3w_1+1)}}\Gamma^2\left(\frac{3(1-w_1)}{2(3w_1+1)}\right)\left[\frac{2(3w_1-1)}{3w_1+1}\right]^2, & \text{if } -\tfrac{1}{3} \leq w_1 < 1
\end{cases},
\end{equation}
\begin{equation}\label{eq:nw1}
    n(w_1) = 
\begin{cases} 
    1 - \frac{3(1-w_1)}{(3w_1+1)}, & \text{if } w_1\ge1 \\
    1 + \frac{3(1-w_1)}{(3w_1+1)}, & \text{if } -\tfrac{1}{3} \leq w_1 < 1
\end{cases},
\end{equation}
and $\Gamma(z) = \int_0^\infty t^{z-1} e^{-t} dt$ is Gamma function. For the two branches presented in Eq.~(\ref{eq:Cw1}) and Eq.~(\ref{eq:nw1}), $ -\frac{1}{3} \leq w_1 < 1 $ corresponds to $ \nu_1 > \frac{1}{2} $, and $ w_1 \geq 1 $ corresponds to $ \nu_1 \leq \frac{1}{2} $. Using this newly obtained general result (Eq.~(\ref{eq:NCkesnw1}) along with Eq.~(\ref{eq:Cw1}) and Eq.~(\ref{eq:nw1})), we can re-derive the two specific cases previously obtained in Ref.~\cite{Li:2024dce}: (1) Setting $ w_1 = \infty $, we recover $ C(\infty) = \frac{1}{2} $, $ n(\infty) = 2 $, and $ N_{22}(\infty) = \frac{1}{2}(k \eta_{\downarrow s})^{-2} $, consistent with Eq.~(60) in Ref.~\cite{Li:2024dce}; (2) Setting $ w_1 = 0 $, we recover $ C(0) = 2 $, $ n(0) = 4 $, and $ N_{22}(0) = 2(k \eta_{\downarrow s})^{-4} $, consistent with Eq.~(70) in Ref.~\cite{Li:2024dce}. This confirms the correctness of the general result.
 
\begin{figure}[htbp]
\centering 
\includegraphics[width=1.0\textwidth]{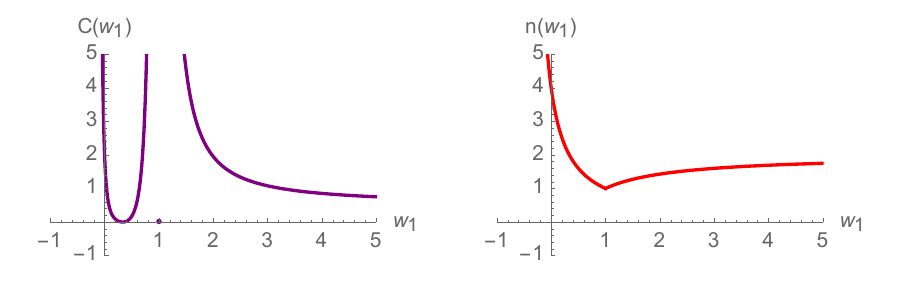}
\caption{\label{fig:Cw1nw1vsw1.pdf} Illustration of $ C(w_1) $ and $ n(w_1) $ using Eq.~(\ref{eq:Cw1}) and Eq.~(\ref{eq:nw1}), respectively. The function $ C(w_1) $ features two poles: 1) $ C(1) = \infty $ at $ w_1 = 1 $, and 2) $ C\left(\frac{1}{3}\right) = 0 $ at $ w_1 = \frac{1}{3} $. These singular points indicate where the analytical approximations break down and will be omitted when plotting the exclusion regions.}
\end{figure}

In Fig.~\ref{fig:Cw1nw1vsw1.pdf}, we illustrate $ C(w_1) $ and $ n(w_1) $ using Eq.~(\ref{eq:Cw1}) and Eq.~(\ref{eq:nw1}), respectively. We note that the general expression for $ C(w_1) $ features two poles: 1) $ C(1) \propto \Gamma^2(0) = \infty $ at $ w_1 = 1 $, and 2) $ C(\tfrac{1}{3}) = 0 $ at $ w_1 = \tfrac{1}{3} $. The reason for these singularities is as follows: at $ w_1 = 1 $, $ \Tilde{\nu}_1 = 0 $, so terms like $ (k \eta_{s\downarrow})^{-\Tilde{\nu}_1} $ become constant and no longer dominate the analytical approximation. Similarly, at $ w_1 = \tfrac{1}{3} $, $ \Tilde{\nu}_1 = \tfrac{1}{2} $, which is equal to $ \Tilde{\nu}_2 = \tfrac{1}{2} $ in our setup (see Eq.~(\ref{eq:ourmodeltnu})), causing the $_{21}$ component of $ M_{1\downarrow \le \tfrac{1}{2}} $ in Eq.~(\ref{eq:mnu1le}) to vanish. Consequently, the leading-order term is lost. Except for these two points, all other values of $ w_1 $ are regular for proceeding with the following analysis. These two singular points are omitted when plotting the exclusion regions in Fig.~\ref{fig:Fullplotasrhos.pdf} and Fig.~\ref{fig:HighResw1asrho.pdf}.

Substituting Eq.~\eqref{eq:NCkesnw1} along with Eq.~(\ref{eq:Cw1}) and Eq.~(\ref{eq:nw1}) into Eq.~\eqref{eq:phfpketa}, and using the definition of bouncing energy scale, $ \rho_{s\downarrow}^{1/4} $, 
\begin{equation}\label{eq:rhosdd}
    \rho_{s\downarrow} \equiv \rho (\eta_s)= 3H^2(\eta_{s\downarrow}) m_\mathrm{pl}^2~,
\end{equation}
we obtain the explicit expression for the PGW spectrum for big bounce cosmology:
\begin{equation}\label{eq:phetakCw1}
    \mathcal{P}_h(\eta_k) = \frac{C(w_1)}{2\pi} \frac{k^{4-n(w_1)}}{H_0^2 \Omega_{\gamma 0} m_\mathrm{pl}^2} \left[H_0^4 \left(\frac{\rho_{s\downarrow}}{\rho_{c0}}\right) \Omega_{\gamma 0}\right]^{\frac{n(w_1)}{4}}.
\end{equation}
Here, the bouncing energy scale, $ \rho_{s\downarrow}^{1/4} $, represents the quasi-highest energy scale of big bounce cosmology, evaluated at the beginning of the post-bounce expansion (Phase IV). To derive Eq.~(\ref{eq:phetakCw1}), we have also used the relations $ H^2(\eta_k) = \frac{k^4}{H_0^2 \Omega_{\gamma 0}} = \left(H_0^2 \Omega_{\gamma 0} \, \eta_k^4\right)^{-1} $ and $ H^2(\eta_{s\downarrow}) = \left(H_0^2 \Omega_{\gamma 0} \, \eta_{s\downarrow}^4\right)^{-1} $ for the radiation-dominated era (Phase IV), the critical energy density today $ \rho_{c0} = 3H_0^2 m_\mathrm{pl}^2 $ with the Hubble parameter today $ H_0 $, and the expression for $ \eta_{s\downarrow} $, $ \eta_{s\downarrow} = \frac{1}{H_0^4}\left(\frac{\rho_{c0}/\Omega_{\gamma 0}}{\rho_{s\downarrow}}\right) $.

\begin{figure}[htbp]
\centering 
\includegraphics[width=0.5\textwidth]{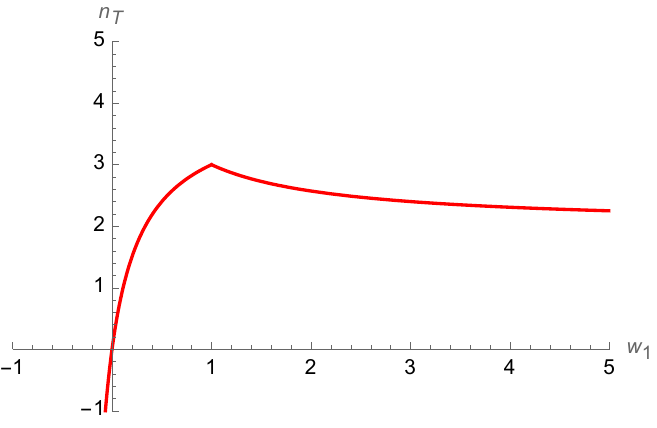}
\caption{\label{fig:ntvsw1.pdf} Illustration of the spectral index of the PGWs $n_T$ as a function of $ w_1 $ for big bounce cosmology by using Eq.~(\ref{eq:nt}) with Eq.~(\ref{eq:nw1}). This illustrates that the PGW spectrum for big bounce cosmology is red-tilted ($ n_T < 0 $) for  $ w_1 < 0 $, blue-tilted ($ n_T > 0 $)  for $ w_1 > 0 $, and scale-invariant ($ n_T = 0 $) for $ w_1 = 0 $.}
\end{figure}

Using Eq.~(\ref{eq:phetakCw1}), we derive the spectral index of the PGWs, $ n_T $, in terms of $ n(w_1) $:
\begin{equation}\label{eq:nt}
    n_T \equiv \frac{d \ln \mathcal{P}_h(\eta_k)}{d \ln k} = 4 - n(w_1)~.
\end{equation}
By combining Eq.~(\ref{eq:nt}) with Eq.~(\ref{eq:nw1}), we plot $ n_T $ as a function of $ w_1 $ in Fig.~\ref{fig:ntvsw1.pdf}. This illustrates that the PGW spectrum for big bounce cosmology is red-tilted ($ n_T < 0 $) for $ n(w_1) > 4 $ (corresponding to $ w_1 < 0 $, as shown in Fig.~\ref{fig:ntvsw1.pdf}), and blue-tilted ($ n_T > 0 $) for $ n(w_1) < 4 $ (corresponding to $ w_1 > 0 $). The case $ n(w_1) = 4 $ ($ w_1 = 0 $) represents the scale-invariant solution for the GW spectrum. This provides new insight into big bounce cosmology, showing that its PGW spectrum (and SGWB spectrum) can be more stringently constrained by low-frequency GW detectors at $ w_1 < 0 $, and high-frequency GW detectors at $ w_1 > 0 $. These constraints will be further reflected in Fig.~\ref{fig:Fullplotasrhos.pdf} and Fig.~\ref{fig:HighResw1asrho.pdf}.

Using $ \mathcal{P}_h(f) = \mathcal{P}_h(\eta_k) |_{k = 2\pi f a_0} $, we obtain the expression for the PGW spectrum in terms of the present-day frequency $ f $:
\begin{equation}\label{eq:phfrhosn}
    \mathcal{P}_h(f) = \frac{C(w_1)}{(2\pi)^{n(w_1)-5}} \frac{f^{4-n(w_1)}}{H_0^2 \Omega_{\gamma 0} m_\mathrm{pl}^2} \left[H_0^4 \left(\frac{\rho_{s\downarrow}}{\rho_{c0}}\right) \Omega_{\gamma 0}\right]^{\frac{n(w_1)}{4}} ~.
\end{equation}
Substituting Eq.~(\ref{eq:phfrhosn}) into Eq.~(\ref{eq:sgwbphf}), we for the first time establish an analytical relation between the bouncing energy scale, $\rho_{s\downarrow}^{1/4}$, and the SGWB, $\Omega_\mathrm{GW}(f)h^2$:
\begin{equation}\label{eq:OmegaGWCnrhos}
    \Omega_\mathrm{GW}(f)h^2 
    =\tfrac{h^2}{24} \left(\tfrac{f_{H_0}}{f_{m_\mathrm{pl}}}\right)^2  \cdot \tfrac{C(w_1)}{(2\pi)^{n(w_1)-5}} \left(\tfrac{f}{f_{H_0}}\right)^{4-n(w_1)} \left[\tfrac{\rho_{s\downarrow}^{1/4}}{\left(\rho_{c0}/ \Omega_{\gamma 0}\right)^{1/4}}\right]^{n(w_1)}  \mathcal{T}_{\mathrm{eq}}(f),
\end{equation}
where $f_{H0}=2.2\times 10^{-18} ~\mathrm{Hz}$ corresponds to $H_0$ and $f_{m_\mathrm{pl}}=3.7\times 10^{42} ~\mathrm{Hz}$ corresponds to the reduced Planck mass $m_\mathrm{pl}$ in natrual units ($\hbar=1$ and $c=1$). 

Additionally, using $a_{s\downarrow} \equiv a(\eta_{s\downarrow}) = \left(\frac{\rho_{c0} \Omega_{\gamma 0}}{\rho_{s\downarrow}}\right)^{1/4}$, we can also relate the bouncing scale factor $a_{s\downarrow}$ to $\Omega_\mathrm{GW}(f)$:
\begin{equation}\label{eq:OmegaGWCnrhosas}
    \Omega_\mathrm{GW}(f)h^2 =\tfrac{h^2}{24} \left(\tfrac{f_{H_0}}{f_{m_\mathrm{pl}}}\right)^2  \cdot \tfrac{C(w_1)}{(2\pi)^{n(w_1)-5}} \left(\tfrac{f}{f_{H_0}}\right)^{4-n(w_1)} \left(a_{s\downarrow}/\sqrt{\Omega_{\gamma 0}}\right)^{-n(w_1)}  \mathcal{T}_{\mathrm{eq}}(f)~.
\end{equation}
These two relations, Eq.~(\ref{eq:OmegaGWCnrhos}) and Eq.~(\ref{eq:OmegaGWCnrhosas}), are equivalent. Physically, $\rho_{s\downarrow}^{1/4}$ and $a_{s\downarrow}$ represent the quasi-highest energy scale and the quasi-minimal size of the Universe in big bounce cosmology. Together, these relations provide a pathway to probe both the energy scale and spatial extent of the early non-singular Universe, while also uncovering new physics through gravitational wave searches.

\section{Systematic Constraints on the Bouncing Energy Scale from Gravitational Waves}

\begin{figure}[htbp]
\centering 
\includegraphics[width=1.0\textwidth]{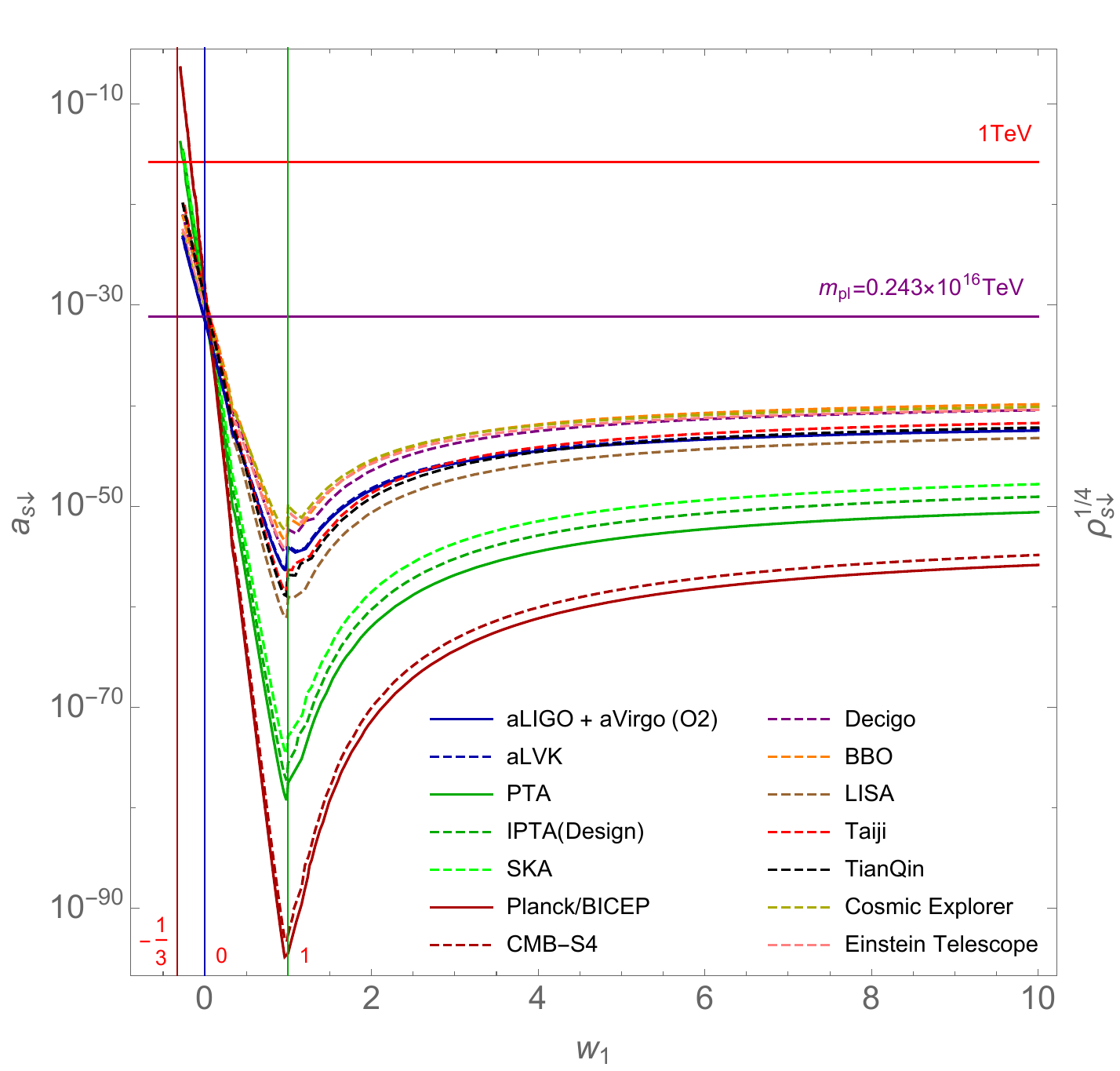}
\caption{\label{fig:Fullplotasrhos.pdf} Systematic GW constraints on the bouncing scale factor $a_{s\downarrow}$ (solid curves for current detectors, dashed for upcoming ones) and corresponding bouncing energy scale $\rho_{s\downarrow}$ (right axis), based on Eq.~(\ref{eq:OmegaGWCnrhosas}) and Eq.~(\ref{eq:OmegaGWCnrhos}). The shaded regions below each curve are excluded by the respective GW constraints. The red and purple horizontal lines mark the conventional bounds of $1~\mathrm{TeV}$ and $m_\mathrm{pl}$, respectively. The vertical lines correspond to $w_1 = -\tfrac{1}{3}$ (red), $w_1 = 0$ (blue), and $w_1 = 1$ (green).}
\end{figure}

In Fig.~\ref{fig:Fullplotasrhos.pdf}, using Eq.~(\ref{eq:OmegaGWCnrhosas}) and Eq.~(\ref{eq:OmegaGWCnrhos}), we present the first systematic gravitational wave  constraint on the bouncing scale factor $ a_{s\downarrow} $ (the quasi-minimal size of the Universe, labeled on the left side) and the corresponding bouncing energy scale $ \rho_{s\downarrow}^{1/4} $ (the quasi-highest temperature of the Universe, labeled on the right side) for big bounce cosmology, based on current observational GW data and forecasts for upcoming detectors. The parameter regions below each curve (corresponding to a smaller bouncing scale factor or larger bouncing energy density) are excluded by the corresponding GW constraints. The two horizontal lines denote $ \rho_{s\downarrow}^{1/4} = 1~\mathrm{TeV} $ (red) and $ \rho_{s\downarrow}^{1/4} = m_\mathrm{pl} = 0.243 \times 10^{16}~\mathrm{TeV} $ (purple), which mark the conventional physically acceptable region for the bouncing energy scale, $ 1~\mathrm{TeV} \lesssim \rho_{s\downarrow}^{1/4} \lesssim m_\mathrm{pl} $. The three vertical lines correspond to $ w_1 = -\tfrac{1}{3} $ (red), $ w_1 = 0 $ (blue), and $ w_1 = 1 $ (green). Specifically, in Fig.~\ref{fig:Fullplotasrhos.pdf}, we include current sensitivities (solid curves)~\cite{Schmitz:2020syl, Annis:2022xgg, Bi:2023tib}: $ \Omega_\mathrm{GW} h^2 = 10^{-9} $ at $ f = 10^{-8}~\mathrm{Hz} $ for PTA (NanoGrav, PPTA, EPTA), $ \Omega_\mathrm{GW} h^2 = 6.9/3.6 \times 10^{-17} $ at $ f = 7.75 \times 10^{-17}~\mathrm{Hz} $ for Planck/BICEP, and $ \Omega_\mathrm{GW} h^2 = 10^{-4} $ at $ f = 3 \times 10~\mathrm{Hz} $ for aLIGO + aVirgo (O2). Additionally, we present forecasts for upcoming detectors (dashed curves), including~\cite{Schmitz:2020syl, Annis:2022xgg, Bi:2023tib}: $ \Omega_\mathrm{GW} h^2 = 5.0 \times 10^{-19} $ at $ f = 7.75 \times 10^{-17}~\mathrm{Hz} $ for CMB-S4; $ 10^{-14} $ at $ f = 2 \times 10^{-9}~\mathrm{Hz} $ for IPTA (design); $ 2 \times 10^{-16} $ at $ f = 2 \times 10^{-9}~\mathrm{Hz} $ for SKA; $ 2 \times 10^{-13} $ at $ f = 1.5 \times 10^{-1}~\mathrm{Hz} $ for DECIGO; $ 3 \times 10^{-14} $ at $ f = 2 \times 10^{-1}~\mathrm{Hz} $ for BBO; $ 5 \times 10^{-12} $ at $ f = 2.5 \times 10^{-3}~\mathrm{Hz} $ for LISA; $ 2.5 \times 10^{-13} $ at $ f = 5 \times 10^{-3}~\mathrm{Hz} $ for TianQin (power-law integrated); $ 0.5 \times 10^{-14} $ at $ f = 2 \times 10^{-3}~\mathrm{Hz} $ for Taiji (power-law integrated); $ 2 \times 10^{-5} $ at $ f = 2 \times 10~\mathrm{Hz} $ for aLIGO + aVirgo + KAGRA (design); $ 9 \times 10^{-10} $ at $ f = 1.5 \times 10~\mathrm{Hz} $ for Cosmic Explorer; and $ 3 \times 10^{-9} $ at $ f = 5~\mathrm{Hz} $ for the Einstein Telescope.

Fig.~\ref{fig:Fullplotasrhos.pdf} shows that current observations from major GW detectors can constrain $ \rho_{s\downarrow}^{1/4} $ across a range from sub-TeV to super-Planck energy scales, depending on the value of $ w_1 $. In particular, low-frequency detectors, such as Planck/BICEP, impose more stringent constraints on $ \rho_{s\downarrow}^{1/4} $ (and $ a_{s\downarrow} $) for $ -\tfrac{1}{3}<w_1 < 0 $, where the PGW spectrum is red-tilted ($ n_T < 0 $, see Fig.~\ref{fig:Cw1nw1vsw1.pdf}). In contrast, high-frequency detectors, like aLIGO + aVirgo (O2), place stronger bounds for $ w_1 > 0 $, where the PGW spectrum is blue-tilted ($ n_T > 0 $). Between these extremes, medium-frequency detectors, such as PTAs, also impose independent constraints on $ \rho_{s\downarrow}^{1/4} $. As a result, constraints from low, medium, and high-frequency detectors converge on the scale-invariant point, $n_T=0$ and $w_1 = 0$, corresponding to the well-known matter-dominated contraction solution in big bounce cosmology.

More specifically, from Fig.~\ref{fig:Fullplotasrhos.pdf}, we find that for $ w_1 \gtrsim 0 $, the current and upcoming GW constraints on $ \rho_{s\downarrow}^{1/4} $ are much larger than $ m_\mathrm{pl} $, implying that current and future GW detectors cannot detect the SGWB induced by PGWs from big bounce cosmology unless the Universe bounces at a super-Planck energy scale, $ \rho_{s\downarrow} \gg m_\mathrm{pl} $. This contrasts with the conventional physically acceptable energy scale for big bounce cosmology, $ 1~\mathrm{TeV} < \rho_{s\downarrow} < m_\mathrm{pl} $. On the other hand, for $ w_1 \lesssim 0 $, current and future GW detectors can exclude a large portion of the parameter space where the Universe's bounce occurs below $ m_\mathrm{pl} $.

\begin{figure}[htbp]
\centering 
\includegraphics[width=1.0\textwidth]{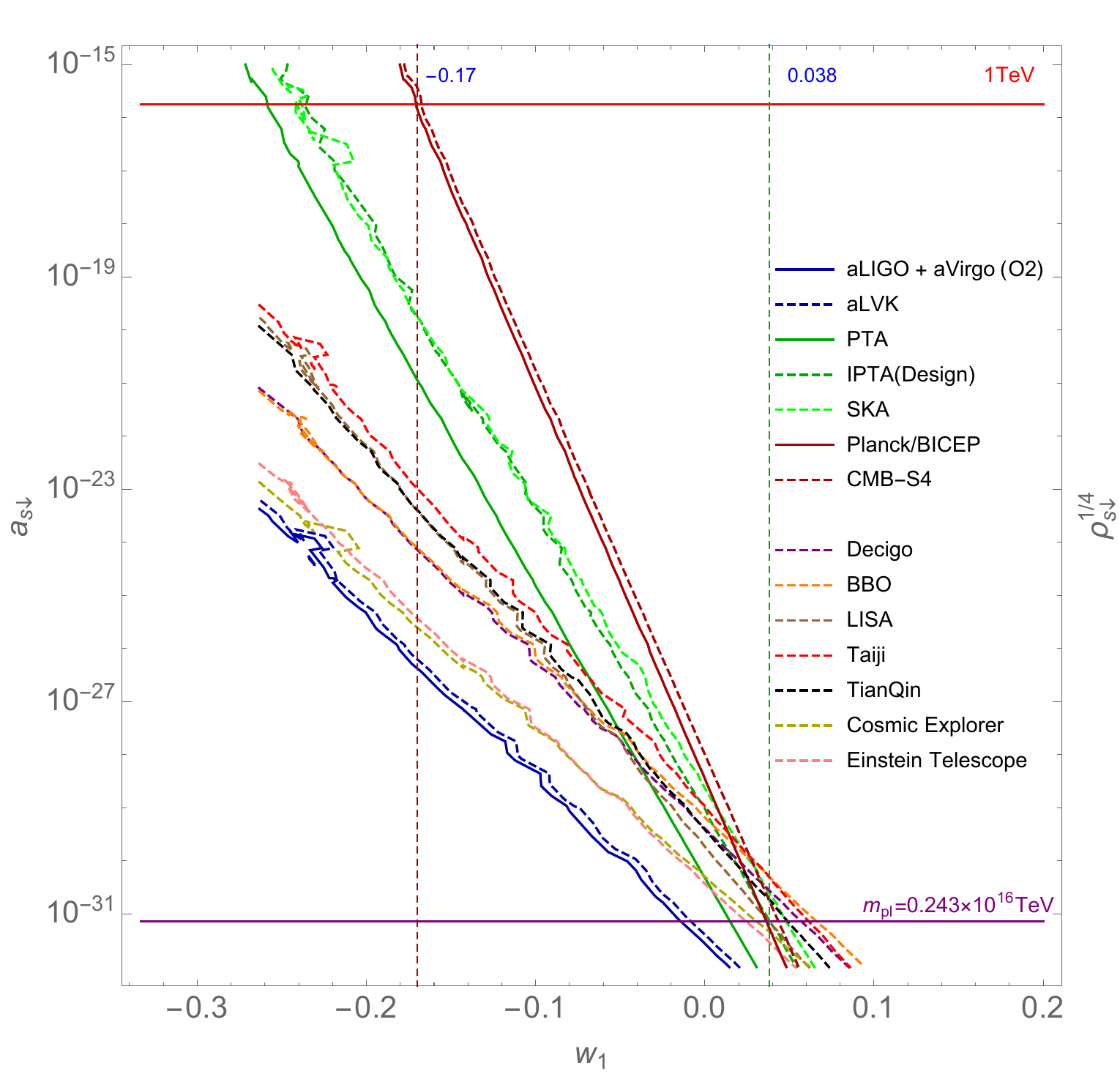}
\caption{\label{fig:HighResw1asrho.pdf} High-resolution plot of Fig.~\ref{fig:Fullplotasrhos.pdf} for $ -\tfrac{1}{3} < w_1 \lesssim 0.2 $. The vertical dashed lines mark $ w_1 = -0.17 $ (red) and $ w_1 = 0.038 $ (green). This plot shows that Planck/BICEP imposes the most stringent constraints on $\rho_{s\downarrow}^{1/4}$ (and $a_{s\downarrow}$) compared to aLIGO + aVirgo (O2) and PTAs in this region. For Planck/BICEP (and CMB-S4), a detectable SGWB can be produced within $ -0.17 < w_1 < 0.038 $, while satisfying $ 1~\mathrm{TeV} < \rho_{s\downarrow}^{1/4} < m_\mathrm{pl} $.} 
\end{figure}

In Fig.~\ref{fig:HighResw1asrho.pdf}, we present a high-resolution plot of Fig.~\ref{fig:Fullplotasrhos.pdf} for the region $ -\tfrac{1}{3} < w_1 \lesssim 0.2 $. The two vertical dashed lines mark $ w_1 = -0.17 $ (red) and $ w_1 = 0.038 $ (green). We find that Planck/BICEP excludes $ w_1 < -0.17 $ under the physical constraint $ \rho_{s\downarrow}^{1/4} > 1~\mathrm{TeV} $, and $ w_1 > 0.038 $ under the constraint $ \rho_{s\downarrow}^{1/4} < m_\mathrm{pl} $. Combining these constraints, we identify a window, $ -0.17 < w_1 < 0.038 $, in which a detectable SGWB can be produced within the physically acceptable range $ 1~\mathrm{TeV} < \rho_{s\downarrow}^{1/4} < m_\mathrm{pl} $. This window excludes nearly all big bounce models, except those with a nearly matter-dominated contraction ($ w_1 \simeq 0 $) (see \cite{Novello:2008ra, Brandenberger:2016vhg, Nojiri:2017ncd, Odintsov:2023weg} and references therein for more details on matter bounce models).

Looking ahead, for $ w_1 < 0 $, the improvement from CMB-S4 is moderate in the parameter space $ (w_1, \rho_{s\downarrow}) $ shown in Fig.~\ref{fig:Fullplotasrhos.pdf} and Fig.~\ref{fig:HighResw1asrho.pdf}, yet potentially decisive: a future CMB-S4 detection of gravitational waves would confirm whether such a signal originates from primordial gravitational waves in big bounce cosmology. If confirmed, both $ w_1 $ and $ \rho_{s\downarrow}^{1/4} $ can be determined from Eq.~(\ref{eq:OmegaGWCnrhos}). On the other hand, for $ w_1 > 0 $, although upcoming detectors (e.g., DECIGO, BBO, LISA, Taiji, TianQin, Cosmic Explorer, and the Einstein Telescope) will significantly improve current limits, they will still fall short of the theoretical constraint at the Planck scale. This suggests that to constrain $ \rho_{s\downarrow}^{1/4} $ for $ w_1 > 0 $ within the physically acceptable energy scale region, high-frequency GW detectors (i.e., $ f \gg 100\,\mathrm{Hz} $) will be needed in the future.

\section{Summary and Prospects}

In this work, we establish for the first time an analytical relation between the bouncing energy scale of big bounce cosmology, $ \rho_{s\downarrow}^{1/4} $, and the SGWB, $ \Omega_\mathrm{GW}(f) h^2 $. By incorporating sensitivities and forecasts from major GW detectors, including Planck/BICEP, PTA, and LIGO/Virgo across low, medium, and high frequencies, we provide the first systematic gravitational wave constraint on $ \rho_{s\downarrow}^{1/4} $. We identify a window, $ -0.17 < w_1 < 0.038 $, where a detectable SGWB can be produced within the physically acceptable range of $ 1~\mathrm{TeV} < \rho_{s\downarrow}^{1/4} < m_\mathrm{pl} $, with the region $ -\tfrac{1}{3} < w_1 < -0.17 $ excluded by $ \rho_{s\downarrow}^{1/4} > 1~\mathrm{TeV} $, and no detectable SGWB generated for $ 0.038 < w_1 < \infty $ under the constraint $ \rho_{s\downarrow}^{1/4} < m_\mathrm{pl} $. This window excludes nearly all big bounce models except those with a nearly matter-dominated contraction ($ w_1 \simeq 0 $), as the matter-dominated contraction model is a scale-invariant solution in the parameter space. Further cosmological applications of this finding warrant additional study.

For $ w_1 < 0 $, the upcoming CMB-S4 detection of gravitational waves could be decisive: if detected, it would confirm whether the signal originates from PGWs in big bounce cosmology and allow for determination of both $ w_1 $ and $ \rho_{s\downarrow}^{1/4} $. However, for $ w_1 > 0 $, while upcoming detectors (e.g., DECIGO, BBO, LISA, Taiji, TianQin, Cosmic Explorer, and the Einstein Telescope) will significantly improve current limits, they will still fall short of the theoretical constraint at the Planck scale due to their relatively low frequencies. This highlights a strong motivation for developing new high-frequency GW detectors (i.e., $ f \gg 100~\mathrm{Hz} $) in the future. In summary, the methodology and results presented in this work pave a promising path for probing the very early non-singular stages of the Universe and uncovering the underlying new physics through current and future gravitational wave searches.

\acknowledgments
C.L. is supported by the NSFC under Grants No.11963005 and No. 11603018, by Yunnan Provincial Foundation under Grants No.202401AT070459, No.2019FY003005, and No.2016 FD006, by Young and Middle-aged Academic and Technical Leaders in Yunnan Province Program, by Yunnan Provincial High level Talent Training Support Plan Youth Top Program, by Yunnan University Donglu Talent Young Scholar, and by the NSFC under Grant No.11847301 and by the Fundamental Research Funds for the Central Universities under Grant No. 2019CDJDWL0005.

%% [A] Recommended: using JHEP.bst file
 \bibliographystyle{JHEP}
 \bibliography{biblio.bib}

%%%%
%%%%

\end{document}